\documentclass[preprint]{aastex}
\shorttitle{Stability in Extrasolar Multiple Planet Systems}
\shortauthors{Rivera \& Haghighipour}

\begin{document}
\title{On the Stability of Test Particles in Extrasolar
Multiple Planet Systems}
\author{Eugenio J.\ Rivera \& Nader Haghighipour}
\affil{Department of Terrestrial Magnetism, 
and NASA Astrobiology Institute,\\
Carnegie Institution of Washington, 
5241 Broad Branch Road, NW, \\
Washington, DC 20015}
\email{rivera@dtm.ciw.edu,nader@dtm.ciw.edu}

\begin{abstract}
We present results of extensive numerical studies of
the stability of non-interacting particles in the planetary
systems of stars $\upsilon$ Andromedae, GJ 876, 47 UMa, and
55 Cancri. We discuss the possibility of the
existence of islands of stability and/or instability at 
different regions in these multi-body systems, and their 
probable correspondence to certain mean-motion resonances.
The results of our study may be applied to questions
concerning the stability of terrestrial planets in these 
systems and also the trapping of particles in resonances 
with the planets. 
\end{abstract}


\section{Introduction}

The number of extrasolar planetary systems
discovered by the precision radial velocity
technique has now risen above 100. Except for
thirteen of these systems, in all others,
the number of detected planets is only one.
It has been noted by \citet{rad+}
that about 50\% of these single-planet systems 
show trends in the residuals to their radial 
velocity fits. This strongly suggests that such systems 
may contain additional companions.

Observations of the present extrasolar multi-planet
systems, on the other hand, have revealed some dynamically 
interesting features of these systems:

i) A few of these systems are in or near mean motion 
resonances (MMR). For instance, the two planets of the 
star GJ 876 are in a 2:1 MMR \citep{GJ876dis,LnC01,RnL01},
the 55 Cancri system ($\rho$ Cnc) has two planets near 
a 3:1 commensurability \citep{55dis,Zhou}, and the two planets in
47 UMa may also be near or in a mean motion resonance 
\citep{47uma2,LCF02,RnH03}.

ii) Some of these systems resemble our Solar 
System. As an example, $\rho$ Cnc has a Jovian-mass
planet with an orbital semi-major axis  
analogous to that of Jupiter \citep{55dis}, and the two 
planets of 47 UMa were announced to have mass- and period-ratios 
resembling those of Jupiter and Saturn \citep{47uma2, RnH03}.

iii) It is also possible that these systems 
may harbor additional low-mass companions. Such bodies
are more amenable to detection with other techniques 
and future space-based missions, such as Darwin, Corot,
Kepler, SIM, and TPF. The precision radial velocity technique
favors the detection of close-in giant planets and 
is currently unable to detect terrestrial planets and/or distant 
gas giants, although continued observations (i.e., extending the time baseline)
can help with the latter insensitivity of this technique.

The increasing number of extrasolar multi-planet systems,
their diversity, and dynamical complexities call
for thorough investigation of their
long-term stability, as well as the dynamical evolution
of possible small bodies that may exist in these systems.
In this paper, we address issues regarding
the stability of non-interacting
bodies in four extrasolar planetary systems 
$\upsilon$ Andromedae, GJ 876, 47 UMa, 
and 55 Cancri. 
Studies of this kind have been done for these and 
a few other extrasolar multi-planet systems. 
However, in this paper, we present results that
have been obtained by integrating these systems using
initial orbital parameters determined by fitting
recent radial velocity data of these stars utilizing a 
newly developed fitting routine. We present the results 
of an extensive study of the stability of test particles 
in these systems, and investigate to what extent these systems can
host additional companions.  We also carry out studies 
to look for regions around these stars where 
low-mass companions, such as terrestrial planets, 
can have long-term stable orbits. Comparison of the
results of our study with those currently available
in the literature is also presented.

The outline of this paper is as follows.  In $\S$\ref{2} 
we discuss the methodology. The subsequent four sections 
present the results of our studies for the four systems 
$\upsilon$ Andromedae, GJ 876, 47 UMa, and 55 Cancri, 
respectively.  In $\S$\ref{fin} we summarize our results 
and discuss their applications. It is necessary to 
mention that any orbital parameter in this study is
an osculating element.

\section{Methodology}
\label{2}

We performed N-body simulations of the planetary systems of
stars $\upsilon$ Andromedae, GJ 876, 47 UMa, and 55 Cancri.  
In order to generate initial conditions for integrating the
orbits of the planets, we performed dynamical (also known 
as N-body or Newtonian) fits to the radial velocity data of 
the four above-mentioned stars using a modified version of 
the Levenberg-Marquardt minimization algorithm 
\citep{NR,LnC01,RnL01}.  

The Levenberg-Marquardt algorithm is a minimization scheme 
that can be used to fit data with a model that depends 
nonlinearly on a set of parameters. We modified the code 
so that the radial velocity data for any star may be fit.
Generally, it is the goodness of the fit, $\chi^2_{\nu}$, 
that is minimized. Close to a minimum, $\chi^2_{\nu}$ is 
assumed to be nearly quadratic in parameters, 
whereas far from it, the algorithm steps toward a minimum
along the gradient of $\chi^2_{\nu}$ taken with respect to 
the parameters, and smoothly varies between these two 
extremes \citep{NR}. 

To fit radial velocity data, the Levenberg-Marquardt scheme 
requires an initial guess for the orbital parameters of each
planet orbiting the star. For the N-body fits performed 
here, the parameters which may be fitted for each planet are 
the mass, $m_{\rm pl}$, the semi-major axis, $a$,
the eccentricity, $e$, the inclination, $i$, the longitude 
of the ascending node, $\Omega$, the argument of periastron, 
$\omega$, and the mean anomaly, $M$. In practice, however,
we fit for $m_{\rm pl}$, $a$, $h$, $k$, and $M$, where
$h=e\sin{\varpi}\mbox{, }k=e\cos{\varpi}\mbox{, 
and }\varpi=\Omega+\omega$. We take the longitudes of the 
ascending nodes in all our fits to be 0$^{\circ}$,
and for most of our fits we assume a 90$^{\circ}$ inclination with respect 
to the plane of the sky. That is, the planets are coplanar,
and their masses have their minimum values.
The best fitted orbital parameters for the 
planetary systems considered here are given in
Tables \ref{upsandparams} -- \ref{55cancriparams}.

Using the parameters determined from the N-body fitting 
routine, we numerically integrated the orbits of a few 
hundred test particles placed in each system. Such
simulations enable us to identify regions where low-mass
companions can have stable orbits. In order to identify 
such a region, one has to integrate the orbit of a 
low-mass object at different distances from the central star. 
We approximate a low-mass body with a test particle
and perform such simulations by
placing non-interacting particles at different distances.
We then perform one simulation with
hundreds of particles at the same time.
Such test particle approximations are similar to those in
\citet[][2001]{RnL00}, \citet{LnR01}, and \citet{MnT03}.

We started all our test particles in the
plane of the planets and on circular orbits with 
respect to the central star. The mean longitudes of all 
particles were taken to be 0$^{\circ}$.  Because different 
initial eccentricities for test particles have considerable 
effects on the zones of stability of the system, 
instead of looking at short term stability for various 
initial eccentricities, we chose to look at long-term 
stability for particles initially only on circular orbits.
No test particle was placed on an orbit that would initially 
cross the orbit of a planet.  

The simulations were performed with the second-order 
mixed variable symplectic (MVS) integrator \citep{WH91} 
in the MERCURY integration package \citep{mercury}.  
This code was modified to include the relativistic 
precession of the longitudes of periastron of all 
bodies in the system \citep{LnR01}.  
In order to investigate if small mutual inclinations 
would have a significant effect on the dynamics of the 
extrasolar systems studied here, we also performed fits 
for fixed mutual inclinations 
of 1$^{\circ}$ and 5$^{\circ}$ between the planets.

\section{$\upsilon$ Andromedae}

Upsilon Andromedae (HD 9826, HR 458, HIP 7513) is an 
F8 V star with a mass of 1.2 $M_{\odot}$ 
(Fischer 2003, private communication).
This star is host to three Jovian-mass planets with periods 
of about 4.6 days, 242 days, and 3.5 years. Using our 
N-body fitting routine, we fitted the radial velocity 
data of 250 observations from JD 2447046.9223 
(September 8, 1987) to JD 2452854.9930 (August 3, 2003).
Table \ref{upsandparams} shows a set of fitted 
astrocentric orbital elements at the epoch JD 2450000.0 
for this system.  The $\sqrt{\chi^2_{\nu}}$ and root 
mean square (RMS) of the differences between the 
observed and modeled radial velocities for this fit 
are 1.58 and 12.94 m\,s$^{-1}$, respectively. Using 
these orbital parameters, we integrated the orbits 
of 410 test particles with semi-major axes ranging 
from 0.06 AU to 10.00 AU. The particles were equally 
spaced every 0.02 AU except for the regions where the
planets made their initial radial excursions. 
The timestep used in this simulation was 0.23 day. 

Figure 1 shows the survival time of a test
particle as a function of its initial semi-major axis.  
The circles represent the planets, and the error bars 
represent the planets' radial excursions due to their
orbital eccentricities. 
Note that since the eccentricity of 
the inner planet is nearly zero, its error bars are 
smaller than the symbol used for this planet.  
As shown in this figure,  
almost all test particles between the outer two 
planets are lost (i.e., ejected from the system, or 
collided with the star) within approximately 3000 years.  
There is a stable region just exterior to the orbit of the 
inner planet where particles survive for at least 10 Myr.  
This is in rough agreement with the results by 
\citet{RnL00} and \citet{LnR01}.  

Figure 1 also shows the locations of several MMRs with the
outer planet.
Except for some long-lived test particles near a few 
of these MMRs, there is a broad region of 
instability, outside the orbit of the outer planet. This region 
extends out to about 7.6 AU, near the location of the 11:2 MMR.  
As shown in \citet{RnL00}, the large size of
this region is likely due to the presence of the 
two massive planets on moderately eccentric orbits.  
This is also in rough agreement with the secular
analysis by Malhotra (2004, in preparation).

A detailed analysis of our simulations reveals that
most of the stable test particles remain in nearly
circular orbits for the duration of the simulation.
Of course, there are a few test particles that
maintained their orbits and also
acquired some eccentricity. For instance, 
five of the six particles from 0.08 AU to 0.18 AU 
acquired a maximum eccentricity less than 0.0509 
over 10 Myr. The one exception was the particle at 0.12 AU,
which became unstable during this time. 

Our simulations also show that for stable test particles 
between 0.20 AU to 0.40 AU, the amplitudes of  
oscillations in both the semi-major axis and eccentricity
rise, and variations in the mean value of the semi-major 
axis start to appear. This implies that the test particles 
closest to the middle planet will not 
survive at longer times.

For distant test particles, one expects small variations 
to appear in both semi-major axis and eccentricity.
One has to note that in the system of 
$\upsilon$ Andromedae, the planets 
exert a significant perturbation on the star.  
From the test particles' point of view, the star 
appears to rapidly revolve around the system's barycenter 
so that the particles' astrocentric velocities have
relatively large variations.  As a result, the 
astrocentric semi-major axes of long-lived and stable 
test particles beyond the orbit of the outermost companion
librate with amplitudes of order 0.2 AU up to 0.5 AU.
We refer the reader to \citet{RnL00} and \citet{LnR01} 
for a discussion of a similar effect on the
semi-major axes of the planets. In general, these amplitudes increase
with increasing distance from the star. For all 
the test particles whose orbital evolution was 
examined here, the libration amplitudes of the barycentric
semi-major axes were not more than about half that of the 
corresponding astrocentric quantities. In a few cases, 
the amplitudes were reduced by 1-2 orders of magnitude. 

The long-lived test particles near MMRs with the outer 
planet warrant further investigation. Just outside the 
1:3 MMR, stable particles experience large oscillations 
in their orbital eccentricities, reaching up into the 
range of 0.2 to 0.3.  
For the duration of the integration, these particles 
are prevented from close approaches with the outer 
planet by the $e$-$\omega$ mechanism 
\citep{MnN84,Glad93,LnR01}. In this mechanism, a test 
particle acquires a low eccentricity when its orbit is
anti-aligned with that of the planet.

An analysis of the critical arguments of test 
particles near the 1:4 MMR suggests that this resonance 
has a significant effect on the stability of these
particles. The eccentricities of these objects either 
undergo gradual growth, or experience jumps to higher 
values.  For the longest lived test particle (7.25 Myr), 
such jumps seem to correlate with the libration or 
circulation of one of its critical arguments,
$\varphi=\lambda_{\rm d}-4\lambda^{\prime}+
\varpi_{\rm d}+2\varpi^{\prime}$.
In this expression
$\lambda_{\rm d}$ and $\varpi_{\rm d}$ represent 
the mean longitude and the longitude of periastron 
of the outer planet, and $\lambda^{\prime}$ and
$\varpi^{\prime}$ correspond to the same quantities 
for the test particle. Larger 
eccentricities tend to occur when this critical argument 
is librating near 0$^{\circ}$. Figure 2 shows the variations of
the eccentricity and $\varphi$ for this particle.

The 1:5 MMR in Figure 1 has a considerable effect on the amplitudes 
of libration of the semi-major axes and eccentricities 
of test particles in its vicinity. Among the seven stable particles in 
this region, the two test particles that are in this 
resonance show the smallest variations in both their 
semi-major axes and eccentricities. The eccentricities 
of the particles in this resonance generally remain below 
0.06 with very regular oscillations, while those outside 
this resonance have typical eccentricities of 0.2 to 0.3 
with irregular oscillations. Figure 3 shows the 
eccentricity versus semi-major axis and the only librating 
critical argument for one of the two particles in
this resonance.

Although most of the remaining five particles near the 1:5 MMR 
have at least one critical argument in 
non-uniform circulation, the variations in their semi-major
axes, and to a lesser extent, their eccentricities, increase 
with distance from the location of this resonance. 
These particles also spend a significant amount of
time in orbits that are anti-aligned with the orbit of
the outer planet. An extension of the simulation to 100 Myr 
shows that the outermost particles, one interior to and 
two exterior to the 1:5 MMR, become unstable after 10 Myr.

\section{GJ 876}
\label{secGJ}

GJ 876 (HIP 113020), an M4 dwarf with a V magnitude 
of 10.1 \citep{perry97}, has a mass of 
0.32 $M_{\odot}$ \citep{M98}. Among currently known 
planet-hosting stars, GJ 876 is the one with the 
lowest stellar mass. This star has two Jovian-mass
companions with periods of about 30 and 61 days, near a 
2:1 MMR. The radial velocity data for GJ 876 was 
originally fitted with a Levenberg-Marquardt
algorithm in which the planets were assumed to be on 
unperturbed Keplerian orbits \citep{GJ876dis}.  
That is, the planet-planet perturbations were not
modeled. However, the relatively small stellar mass, 
the large planetary masses, and their close spacing  
cause significant perturbations between the planets of 
this system. That makes the N-body fitting routine, 
as explained in $\S$\ref{2}, a more appropriate scheme
for fitting the radial velocity data of this star.

To obtain the orbital parameters of the two planets,
we carried out three independent fits, 
corresponding to a total of 106 observations from 
JD 2450602.0931 (June 2, 1997) to JD 2452851.0567 
(July 30, 2003), to the radial velocity data of GJ 876.
In the first fit, 
we assumed that the system was edge-on, as viewed from Earth.  
For the second fit, we adopted an inclination
of 84$^{\circ}$ with respect to the plane of the 
sky for both planetary orbits. This is equal to the 
astrometrically determined inclination of the outer
companion as reported by \citet{astro}. In the third fit,  
we varied the orbital inclinations of both planets by
the same amount, keeping their orbits coplanar, and
only fitted for the remaining 10 parameters that were 
listed in $\S$\ref{2}. We then performed the
fitting procedure for various values of this inclination
until a minimum in $\sqrt{\chi^2_{\nu}}$ was obtained
\citep[see][for more details on this procedure]{RnL01}. 
Figure 4 shows $\sqrt{\chi^2_{\nu}}$ versus $i$
for this fitting.
As shown here, a minimum is reached where 
$i \sim 39.6^{\circ}$. It is important to mention that $\sqrt{\chi^2_{\nu}}$
is not sensitive to values of $i>35^{\circ}$. 
Thus, other values of the inclination larger than
35$^{\circ}$ are nearly as good as the best one.

Table \ref{GJ876params} shows the fitted astrocentric 
orbital parameters at epoch JD 2451310.0 for the three 
fits mentioned above.  As shown in this table,
the $\sqrt{\chi^2_{\nu}}$ is the same for all three fits  
implying that these fits are statistically identical. 
We performed numerical simulations of the orbital 
stability of test particles for each fit.  
The timesteps for all three simulations were 1.0
day. For the first fit, we considered 318 test 
particles placed within the range of 0.160 AU to 
0.800 AU at equal intervals of 0.002 AU. As discussed in
$\S$\ref{2}, no test particle was placed on an orbit 
that would initially cross the orbits of the planets.  
Since the semi-major axes and eccentricities of the
planets, obtained from the second fit, are very similar 
to those of the first fit, we considered an identical 
set of test particles for the second simulation.  
Figure 5 shows the results of these simulations.
Since the initial conditions for these two 
simulations are nearly identical, the results are 
qualitatively the same. These results are also similar to
those of \citet{RnL01}, in which the authors considered a fit
with $i=37.0^{\circ}$.

As shown in Figure 5, in both simulations, 
the region between the two planets is cleared
out in less than 530 years. This figure also shows that,
out to about 0.294 AU from the central star,
no test particle survives for more than 530 years.
Unlike \citet{RnL01}, each of these two simulations 
contains a region extending from 0.294 AU to 0.310 AU, 
in the vicinity of the 4:7 MMR with the 
outer planet, where 5 or 6 test particles survive 
beyond $10^4$ years. All but two of these particles are 
lost within 250000 years from the start of the 
simulations. Of these two long-lived particles, 
one survives for 6.3 Myr.

Figure 5 also shows a region of instability 
extending from 0.312 AU to 0.336 AU, near the 
1:2 MMR with the outer planet.  
All particles in this region are lost 
within $10^4$ years.  Beyond this region, from 
0.336 AU to 0.428 AU, except for one particle at 
0.338 AU, which survives for only 2.85 Myr, 
all other particles are stable for the entire 
duration of the simulations.  The last set of unstable
particles is between 0.428 AU and 0.438 AU,
around the 1:3 MMR with the outer planet.
In this region, for each simulation, all but 
one test particle at 0.436 AU, are lost in about 1 Myr.
Continuations of these simulations beyond 10 Myr 
show that only two test particles are lost in each 
simulation; the stable one in the 1:3 MMR region,
and one at 0.382 AU, near the location of the 2:5 
commensurability with the outer planet.

In these simulations, the region near the 1:3 MMR 
with the outer planet is probably the most dynamically 
interesting. The eccentricities of unstable particles 
in this region undergo sudden jumps. Inspection of the
three critical arguments for the 1:3 MMR for all these 
unstable particles suggests that at least one of their 
critical arguments repeatedly changes
from circulation to large amplitude ($\sim 180^{\circ}$) 
libration. This is a characteristic of a particle that 
repeatedly goes in and out of resonance. Figure 6 
shows the time evolution of the eccentricity and the 
critical argument
$\varphi=\lambda_{\rm b}-3\lambda^{\prime}+2\varpi_{\rm b}$ 
of one of these particles. In this formula, $\lambda_{\rm b}$, 
and $\varpi_{\rm b}$ represent the mean longitude and the 
longitude of periastron of the outer planet, and 
$\lambda^{\prime}$ corresponds to the mean longitude 
of the test particle. The plot of the critical argument $\varphi$ 
shows that the range 0$^{\circ}$ to 360$^{\circ}$ is not 
uniformly filled at all times. This indicates that the 
critical argument is not executing pure circulation.

The most stable test particles in the region near 
the 1:3 MMR with the outer planet
have small eccentricities. The amplitudes 
of librations of the semi-major axes and eccentricities 
of these particles
are relatively small, and all their three critical 
arguments circulate. Figures 7 and 8
show the graphs of the eccentricity versus semi-major axis, 
and the time evolution of all three critical arguments
of one such particle, respectively.
In both simulations, the continuations of the 
integrations to longer times indicate that the 
eccentricity of the single stable particle in that region 
undergoes a sudden jump, and one of 
its critical arguments starts librating. 
Subsequently, this particle is lost.

In the simulations above, one stable particle 
at the outer edge of the region near the 1:3 MMR 
with the outer planet shows a different evolution 
in one of its critical arguments, 
$\varphi=\lambda_{\rm b}-3\lambda^{\prime}+2\varpi_{\rm b}$.  
While the eccentricity of this particle remains low, 
its critical argument actually librates about 
180$^{\circ}$ with an amplitude of $\sim$ 100$^{\circ}$.
Figure 9 shows the eccentricity of this particle
versus its semi-major axis, and its librating
critical argument for the  simulation with
$i=90^{\circ}$.  This evolution of the critical argument 
was also observed in the continuations of both 
simulations to longer times.

Figure 10 shows the survival times of test 
particles in the simulation based on the third set of 
fitted planetary parameters (i.e., $i \sim 39.6^{\circ}$).  
Since the orbital eccentricities of the planets obtained
from the third fit were somewhat larger than in the 
first two, the set of test particles for the third 
simulation was slightly different. A total of 
nine test particles from the first two sets were excluded 
due to their crossing 
orbits with the planets.  The rather larger masses and 
eccentricities obtained from the third fit also have 
a significant effect on the long-term stability of 
particles in and near several MMRs with the outer
planet. In particular, compared with the previous 
two fits, the region between the 1:2 and 1:3 MMRs 
and also the region around the 1:3 MMR are far 
less stable. 

Figure 10 also shows that in this
simulation, there is still one test particle in 
the instability region around the 1:3 MMR which behaves
like most stable test particles in the other 
two simulations. There is also 
a test particle out as far as the location of the 1:4 MMR
with the outer planet that
was lost in less than 1 Myr. This result is in 
agreement with the general finding of \citet{LnD04} 
who indicated that the strength of the MMRs increases
with the planetary masses and eccentricities 
\citep[see also][]{MD,partials}. An extension of this 
simulation to larger times indicates that seven more test
particles are lost after 10 Myr.  All but one of these 
particles are at the edges of the three islands of 
stability closest to the outer planet.

\section{47 UMa}

The star 47 UMa (HD 95128, HR 4277, HIP 53721) has 
a spectral type of G0 V. The mass of this star is 
1.03 $M_{\odot}$ and it is host to two planets.
Initial fits to the radial velocity data of this star 
by \citet{47uma2} indicated  minimum masses of
2.54 $M_{\rm Jup}$ and 0.76 $M_{\rm Jup}$, and periods 
of 1089 days and 2594 days, for the two planets
of this system, respectively.
It was also shown by \citet{47uma2} that there are 
large uncertainties in the orbital parameters of the 
outer planet, particularly, in its eccentricity.
This can be attributed to the short baseline of 
observations (13 years at that time) compared to the 
orbital period of the outer planet.  Because of this
short baseline, the N-body fitting routine discussed in $\S$2
is incapable of fitting the parameters for this system
properly\footnote{See Rivera \& Lissauer 2001 for a 
presentation of some problems with N-body fitting routines 
and how one can try to address these problems.}.
Several sets of parameters produce equally good fits. 
To overcome this difficulty, we iteratively fit for the 
parameters of one planet while keeping the parameters
of the other planet constant.
We utilized this iterative fitting procedure for 128 
observations from JD 2446959.7372 (June 13, 1987) to 
JD 2452834.6980 (July 14, 2003), which includes
thirty-seven new observations in addition to those of
\citet{47uma2}.  Our results indicate that the most 
stable system with the lowest $\sqrt{\chi^2_{\nu}}$, 
is obtained by holding the parameters of the
inner planet constant and fitting for those of the outer 
planet. Table \ref{47umaparams} shows the fitted 
astrocentric orbital parameters of this system at 
epoch JD 2449900.0.

We numerically integrated the orbits of 469 test particles 
in the 47 UMa planetary system. The timestep for 
this simulation was 10.0 days, and particles were placed 
at equal intervals of 0.02 AU, with semi-major axes 
ranging from 0.44 AU to 10.00 AU. Figure 11
shows the survival times of these test particles as a 
function of their initial semi-major axes. The resonances 
marked inside the orbit of the inner planet are internal 
MMRs with the inner planet, and those outside the orbit 
of the outer planet are external MMRs with the outer 
planet. As shown here, the region from the 5:3 MMR 
with the inner planet ($\sim$ 1.48 AU) to the 2:3 MMR
with the outer planet ($\sim$ 5.85 AU) is cleared out 
in approximately 10$^5$ years. An exception to this 
is observed for several test particles that are 
temporarily trapped in a 1:1 MMR with the outer planet.  
This is not unexpected since the orbit of the outer 
planet is nearly circular \citep{Danby}.

Figure 11 also shows that near the locations of 
several other MMRs, test particles can survive for
extended times. Our results indicate that 219 particles 
are lost by 10 Myr from the beginning of the simulation.  
An extension of the integration to 100 Myr shows that a 
significant number of the remaining particles, 
including those around resonances shown in Figure 11, 
are lost in less than 50 Myr. 
Among the particles that are lost after 10 Myr 
are the ones in the vicinity of the 1:3 MMR with the 
outer planet.  Similar to the unstable test particles 
near the 1:3 MMR with the outer planet in the GJ 876 
system, the eccentricities of these particles undergo 
sudden jumps indicating transitions from circulation 
to large amplitude libration for at least one of their 
critical arguments. Also, only two new instability 
regions appear around 7.94 AU and 8.67 AU as three 
test particles are lost from these areas in 43 Myr.

An interesting outcome of the extended simulation is 
the stability of test particles in the habitable 
region of 47 UMa. Similar to the results reported
by \citet{NMC02}, our simulation shows that test particles
are stable in this region for 100 Myr.  
Also, as opposed to \citet{RnH03}, 
who have shown that a test particle at 1 AU, near the 3:1 MMR 
with the inner planet, could not be in a stable orbit beyond 
$3\times10^5$ years, our simulation indicates that such a 
particle would be stable for at least 100 Myr.  
The differences arise from the different initial conditions
used in this study.
Figure 12 shows the three critical arguments 
of the test particle at 1 AU, near the 3:1 MMR, over 100 Myr.
Since all three critical arguments of this particle 
circulate (although not uniformly), this particle appears 
to be stable for this length of time. It is, however, 
necessary to note that a small change in the 
position of the inner planet or the test particle could 
change this situation.

Figure 13 shows the graphs of eccentricities versus 
semi-major axes for particles in the region of 0.9 AU to 1.5 AU
in the 47 UMa planetary system.
As shown here, this region is populated
by resonances.  Most of these resonances are located 
beyond 1 AU. This dynamically complex 
region has also been discussed by \citet{LCF02}, 
\citet{G02}, and \citet{Asghari}, who studied 
both mean motion and secular resonances.
However, significant differences exist between our results 
and the results of the above-mentioned studies. These 
differences are most likely due to
the different planetary orbital parameters used here.  
Since the semi-major axis of the outer planet used in this study
is very different from that used by these authors, 
the positions of secular resonances
in our simulation are entirely different.  
This has a significant effect on
test particle stability \citep{LnD04}.

The most noticeable feature of test particles in 
this region is that most of those with semi-major axes 
below 1 AU attain maximum eccentricities no more than 0.05
over 100 Myr. Figure 13 shows the eccentricities 
of these particles versus their semi-major axes. 
For a comparison, the widths of the 2:1 and 3:1 MMRs with
the inner planet have also been shown \citep{MD,partials}.
The small width of the 3:1 resonance for small eccentricities
strongly indicates that the extreme sensitivity to 
initial conditions probably contributes to the 
differences seen in the behavior of test particles near 1 AU
in the 47 UMa system \citep{LCF02,RnH03}.  
This suggests that for a slightly different initial 
semi-major axis within the width of the resonance,
a test particle might be unstable, which
would more closely agree with \citet{LCF02} and 
\citet{G02}. Another interesting feature of particles
in the region from 0.90 AU to 1.50 AU is that 28 out of 31 
test particles in this region spend at least 60\% 
of their lifetimes in orbits which are within 90$^{\circ}$ of
being aligned with the orbit of the inner planet.

Figure 14 shows a test particle at 
1.32 AU. This particle spends most of its time in the 2:1 MMR 
with the inner planet. This configuration raises the
eccentricity of this particle, which generally
results in orbital instability. However, 
this particle  survives for 100 Myr. Figure 14 
also shows the eccentricity of this particle versus 
its critical argument,
$\varphi=\lambda^{\prime}-2\lambda_{\rm b}+\varpi^{\prime}$.
In this formula, $\lambda_{\rm b}$ is the mean longitude
of the inner planet, and 
$\lambda^{\prime}$ and $\varpi^{\prime}$ represent 
the mean longitude, and the longitude of periastron
of the test particle. As shown here, for most of the 
simulation, this critical argument librates. However,
since $\varphi$ circulates non-uniformly for several nearby
stable test particles at smaller semi-major axes, 
this 2:1 MMR appears to have considerable influence 
on the orbital evolution of such particles. The critical arguments
of these objects
spend more time near 0$^{\circ}$ than at 180$^{\circ}$.

\section{55 Cancri}

The main sequence star 55 Cancri (HD 75732, HR 3522, HIP 43587) 
has spectral type G8 with a mass of 0.95 $M_{\odot}$.  
This star is host to three planets with minimum 
masses of about 0.8 $M_{\rm Jup}$, 0.2 $M_{\rm Jup}$, and
4 $M_{\rm Jup}$, and periods of about 14.65 days, 
44 days, and 5360 days, respectively.
We performed N-body fits to the radial velocity data  
of 170 observations of this star carried out from 
JD 2447578.7300 (February 21, 1989) to JD 2452737.7040
(April 8, 2003).
Table \ref{55cancriparams} shows the astrocentric orbital 
parameters of the planets at epoch JD 2450165.0.
The values of $\sqrt{\chi^2_{\nu}}$ and the RMS for 
this fit are 2.34 and 8.79 m\,s$^{-1}$, respectively.
Using these parameters, we numerically integrated 
the orbits of 455 test particles placed in this system 
from 0.14 AU to 12.00 AU, at equal intervals of
0.02 AU. The timestep for this simulation was 0.5 day.  
Figure 15 shows the survival times of these 
test particles as a function of their initial
semi-major axes. The MMRs marked in this figure are 
the internal and external MMRs with the outer planet 
for the ones inside and outside its orbit, respectively.  
The results here resemble those in \citet{RnH03} 
except that because the orbital parameters for the 
outer planet are quite different, a different set of 
MMRs affect the stability of test particles in the regions
of 2 AU to 3 AU, and 10 AU to 12 AU.

As shown in Figure 15, there is a small region, 
extending from 0.30 AU to 0.60 AU, just outside the orbit
of the middle planet, where most test particles 
are unstable. The survival time of test particles in this region
generally rises with increasing distance 
from the planet. The last unstable test particle 
in this region is near the outer 1:4 MMR with 
the middle planet. The simulation also indicates a broad region of 
stability from 0.62 AU to 2.36 AU.
Such a stable region has also been reported 
by \citet{55dis} who also showed that a terrestrial-mass
planet at a distance of 1 AU from the central star would be stable.  
\citet{BnR04} have also found a similar broad stable 
region in this system. 
These authors find that
the stability of particles is not 
affected significantly by starting them on orbits 
with eccentricities up to 0.25.

Mean motion resonances affect the stability of 
test particles depending on their order and also 
the locations of the particles. In this simulation, 
the 7:2 MMR affects the stability of the
test particle at 2.38 AU by pumping up its 
eccentricity until it is lost in 47000 years.  
Gradual drifts were also observed in the semi-major 
axes of the two nearest neighbors of this particle. 
Such drifts may cause these particles to become 
unstable on timescales beyond 100 Myr. The 3:1 and 
8:3 MMRs both destabilize a few test particles in 
their vicinity. The test particles in this small 
region that survive at least 10 Myr mark the 
boundary between the broad stable region and a
broad unstable region in this system.
The broad unstable region extends out to 
$\sim$ 10.1 AU, the location of the 2:5 MMR with 
the outer planet. The last significant region of 
unstable test particles is just outside the 3:8 MMR 
with the outer planet at $\sim$ 10.5 AU.

An extension of this simulation to 100 Myr 
resulted in the loss of an additional 12 particles. 
Among these particles, six were at the edge of an
island of stability near 3 MMRs with the outer planet.  
Our extended simulation also showed that three test 
particles at the inner boundary of the broad region
of stability were lost, and that three new unstable 
regions appeared at 1.66 AU, 10.82 AU, and 11.40 AU.  
The latter region corresponds to the 1:3 MMR with the
outer planet. The test particles around this 
resonance show three types of behaviors that 
are similar to those observed for test particles
near the 1:3 MMR with the outer 
planet of GJ 876.
In correlation with an increase in their 
eccentricities, two stable particles in this region 
showed an increase in the libration amplitudes of one 
of their critical arguments. Also, the stable 
particle at 11.42 AU in this region had two critical 
arguments turning from circulation to libration, 
with a corresponding jump in its eccentricity.  
Based on the results for GJ 876, it is likely that this region
is unstable on longer times beyond 100 Myr.
This is the same mechanism that makes some resonances
(e.g., 3:1) unstable in the asteroid belt.

Between the two outer planets of this system,
a stable region exists from 0.7 AU to 1.3 AU, 
which includes the star's habitable zone \citep{MnT03}.
To ensure that our study would also include an  analysis of
the stability of test particles in this region,
we took a closer look at the stability of particles 
in the region from 0.64 AU to 2.90 AU.  Similar to
the situation in 47 UMa, this region is populated 
by resonances with the middle and outer planets.  
The most easily discerned feature of 
the test particles in this region is that their 
maximum eccentricities rise almost monotonically 
from about 0.03 at 0.76 AU to 0.4 at 2.90 AU.  
The amplitudes of libration of the semi-major
axes of these particles also increase in this range.  
Figure 16 shows the eccentricity versus 
semi-major axis of all test particles with initial 
semi-major axes between 0.64 AU and 1.48 AU. As shown 
in this figure, only the particles closest to 
the middle planet appear to be strongly affected by
its presence, although all the test particles 
in this figure are stable for 100 Myr. Figure 17 
shows the same quantities for test particles with
initial orbital radii between 1.50 AU and 2.90 AU.  
In this figure, the indicated MMRs are internal resonances
with the outer planet.  The width of the 3:1 MMR is
also shown here. The MMRs with the outer planet 
of the form $n:1$ have a significant effect on the 
eccentricities of particles at their nominal locations.

\section{Summary and Discussion}
\label{fin}

We have studied the stability of non-interacting 
particles in the multi-planet extrasolar systems of
$\upsilon$ Andromedae, GJ 876, 47 UMa, and 55 Cancri.
We identified regions of stability of these systems where test 
particles maintain their orbits for 100 Myr without 
being ejected from the system or colliding with the central
star. Although examples of secular resonances, in which the 
lines of apsides remain closely aligned, are also present in a 
few extrasolar systems \citep{Chiang01,Chiang02,Malhotra}, 
we restricted ourselves to study of the effects of mean
motion resonances on the dynamical evolution of test particles
in these planetary systems.

The simulations presented in this study were 
initially run for 10 Myr and were subsequently 
extended to 100 Myr. The results indicate that particles 
near MMRs with planets and at the edges of the 
zones of stability could be lost very slowly over 
extended times.  Most particles placed in regions 
between planets became unstable on short timescales
($<10^5$ years).  Three exceptions to this were 
the region just outside the orbit of the inner planet 
in $\upsilon$ Andromedae, some captured Trojans in 47 UMa, 
and also the region between the middle and outer planets 
in 55 Cancri.

Among all the mean motion resonances studied 
in this paper, the external 1:3 MMR played a more significant 
role in the dynamical evolution of test particles.
For this resonance, a test particle is stable 
either when all its three critical arguments
circulate for the length of the simulations, or when 
only one of its critical arguments librates for 
that duration of time. Instability generally arises when 
at least one critical argument undergoes a transition 
between circulation to libration. Sudden jumps in 
eccentricity accompany such transitions. This behavior 
has also been observed in simulations of test
particles near Jupiter's internal 3:1 MMR, the location 
of a Kirkwood gap \citep{Wis83}.

Studies such as the one presented here are the first
step in attempting to identify 
extrasolar planetary systems that may harbor 
terrestrial planets. Terrestrial planets are on 
average about 2 orders of magnitude less massive 
than Jupiter-like planets.  Unless two terrestrial
planets are close together (e.g., Earth-Venus), or
are involved in a secular resonance with giant planets
(e.g., Mercury-Venus-Jupiter, cf.\ Laskar 1994, 1997),
the perturbative effects of such bodies on the 
dynamics of the system is so small that to a good 
approximation, they can be neglected.  This enables
one to consider terrestrial planets as test particles.  
To search for regions where a terrestrial planet 
can have a stable orbit, one has to run simulations 
for different values of the orbital parameters of 
such a body.  In this study we accomplished this by 
considering a battery of test particles, systematically
placed at different semi-major axes, to represent 
a terrestrial planet at those locations. Such
studies have also been done by other authors. However,
this study appears to be one of the first in which 
the planetary orbital parameters were determined 
by performing N-body fits to radial velocity data.

In general, most of our results agree with 
previous studies. As \citet{JnS03}, \citet{JSC01},
and \citet{MnT03} 
have found, it is unlikely that $\upsilon$ Andromedae 
and GJ 876 harbor a terrestrial planet in a stable 
orbit in their respective habitable zones.  
However, for both 47 UMa and 55 Cancri, the existence 
of such a terrestrial planet is not impossible.  
It is necessary to emphasize that, if significant giant 
planet migration occurs during the formation of the two 
latter systems, it will be difficult for a terrestrial 
planet to maintain a stable orbit for the duration of the 
formation \citep{LCF02,55dis}. As the giant planet 
migrates, the locations of the mean motion and secular 
resonances also vary. As a result, either the terrestrial 
planet will be removed from the system, or the material 
that needs to go into the making of such a planet will be 
depleted by the passage of mean motion and secular
resonances.

Our simulations also indicate that the extrasolar
multi-planet systems studied here could
harbor more planets in one of the following four 
regions: 1) close-in orbits, even
if the system already has a Jovian-mass planet 
very close to the star as in $\upsilon$ Andromedae, 
2) distant orbits, 3) orbits between widely spaced
planets, and 4) mean motion and secular resonance-protected 
orbits, where a small planet could be protected 
from close encounters with its Jovian neighbors.

Systems like 47 UMa and 55 Cancri, which contain 
significant regions of stability inside the orbit 
of the outermost planet, could also harbor asteroids 
in addition to, or instead of additional planets.
The asteroid belt in our Solar System has two 
properties that are also apparent in the zones of 
stability in the extrasolar planetary systems 
studied here; there are gaps at the locations of 
MMRs with Jupiter, and also the belt is very slowly 
being eroded \citep{NM98}. This may also happen in extrasolar 
asteroid belts. Similar to the Solar System,
collisions among these extrasolar asteroids 
would produce dust. This asteroidal dust could be 
detected by ground-based interferometry and also by TPF
and Darwin. Resolving 
structures in this dusty environment, over scales 
of a few AU from the star, would help in constraining the
parameters for the Jovian and possibly other nearby planets.

It is important to emphasize that the work presented here is
based on recent radial velocity data. Also,
as mentioned before, some of the orbital parameters of the
planetary systems studied here are not strongly constrained.
For instance, a range of values for a planet's orbital inclination
can result in equally good fitted orbital parameters.  
As the data acquisition techniques are modified,
instruments are improved, and more observational data are
gathered, it is likely that the best fitted orbital
parameters will change. How significant
these changes will be, and how they will help in constraining
the orbital parameters depend partly on the length of time for
which a system has been observed. With the future data,
it will be vitally important to
conduct studies similar to the one presented here
to refine the orbital parameters of planetary systems.
For instance, if future data indicate that the masses and
eccentricities of planets in some systems are larger than the current
best values, based on this work and similar
studies, the regions around MMRs would possibly become 
less stable. For now, however, studies
such as the one presented here
can be used to predict where possible additional planets
may exist in extrasolar planetary systems, based on all the recent data.   
\acknowledgments
We are grateful to Alan Boss and John Chambers for
critically reading the original manuscript, and for
their helpful comments and suggestions. 
We thank Debra Fischer, Geoff Marcy, and Paul Butler 
for providing the radial velocity observations and fits 
for the systems studied in this work. We also
appreciate Alycia Weinberger for insightful discussions.  
This research is supported by the NASA Astrobiology 
Institute under Cooperative Agreement NCC2-1056 for 
E. J. R. and N. H., and also by the NASA Origins of Solar Systems
Program under Grant NAG5-11569 for N. H.

\clearpage

\setcounter{table}{0}

\begin{deluxetable}{lccc}
\tablewidth{0pt}
\tablecaption{Astrocentric orbital parameters for 
the three planets orbiting
$\upsilon$ Andromedae. \label{upsandparams}}
\tablehead{\colhead{Parameter} & \colhead{inner} 
& \colhead{middle} & \colhead{outer}}
\startdata
$m_{\rm pl}$ ($M_{\rm Jup}$) & 0.65 & 1.80 & 3.59\\
$a$ (AU) & 0.0577 & 0.807 & 3.43\\
$e$ & 0.0112 & 0.276 & 0.268\\
$\omega$ (deg) & 53.8 & 249.9 & 261.2\\
$M$ (deg) & 215.4 & 119.4 & 337.6\\
\enddata
\end{deluxetable}

\clearpage

\begin{deluxetable}{lcc}
\tablewidth{0pt}
\tablecaption{Astrocentric orbital parameters for 
the two planets orbiting GJ 876. \label{GJ876params}}
\tablehead{\colhead{Parameter} & \colhead{inner} & \colhead{outer}}
\startdata
\cutinhead{\parbox{160pt}{\hspace*{\fill}$i=90^{\circ}$ 
\hspace*{\fill}\\$\sqrt{\chi^2_{\nu}}=1.60$, RMS = 7.77 m\,s$^{-1}$}}
$m_{\rm pl}$ ($M_{\rm Jup}$) & 0.59 & 1.88\\
$a$ (AU) & 0.130 & 0.208\\
$e$ & 0.224 & 0.0152\\
$\omega$ (deg) & 330.1 & 318.6\\
$M$ (deg) & 164.8 & 271.6\\
\cutinhead{\parbox{160pt}{\hspace*{\fill}$i=84^{\circ}$ 
\hspace*{\fill}\\$\sqrt{\chi^2_{\nu}}=1.60$, RMS = 7.77 m\,s$^{-1}$}}
$m_{\rm pl}$ ($M_{\rm Jup}$) & 0.60 & 1.89\\
$a$ (AU) & 0.130 & 0.208\\
$e$ & 0.224 & 0.0156\\
$\omega$ (deg) & 330.0 & 318.6\\
$M$ (deg) & 164.9 & 271.7\\
\cutinhead{\parbox{160pt}{\hspace*{\fill}$i\approx39.6^{\circ}$ 
\hspace*{\fill}\\$\sqrt{\chi^2_{\nu}}=1.60$, RMS = 7.82 m\,s$^{-1}$}}
$m_{\rm pl}$ ($M_{\rm Jup}$) & 0.90 & 2.96\\
$a$ (AU) & 0.130 & 0.209\\
$e$ & 0.288 & 0.0379\\
$\omega$ (deg) & 329.6 & 302.2\\
$M$ (deg) & 165.2 & 282.1\\
\enddata
\end{deluxetable}

\clearpage

\begin{deluxetable}{lcc}
\tablewidth{0pt}
\tablecaption{Astrocentric orbital parameters for 
the two planets orbiting 47 UMa. 
($\sqrt{\chi^2_{\nu}}=1.55$, RMS = 10.01 m\,s$^{-1}$)
\label{47umaparams}}
\tablehead{\colhead{Parameter} & \colhead{inner} & \colhead{outer}}
\startdata
$m_{\rm pl}$ ($M_{\rm Jup}$) & 2.72 & 1.02\\
$a$ (AU) & 2.09 & 4.47\\
$e$ & 0.0466 & $5.97\times10^{-6}$\\
$\omega$ (deg) & 108.2 & 328.6\\
$M$ (deg) & 270.6 & 330.7\\
\enddata
\end{deluxetable}

\clearpage

\begin{deluxetable}{lccc}
\tablewidth{0pt}
\tablecaption{Astrocentric orbital parameters for the 
three planets orbiting 55 Cancri. \label{55cancriparams}}
\tablehead{\colhead{Parameter} & \colhead{inner} & 
\colhead{middle} & \colhead{outer}}
\startdata
$m_{\rm pl}$ ($M_{\rm Jup}$) & 0.82 & 0.18 & 3.74\\
$a$ (AU) & 0.115 & 0.240 & 5.49\\
$e$ & 0.0657 & 0.201 & 0.244\\
$\omega$ (deg) & 122.4 & 33.8 & 196.5\\
$M$ (deg) & 34.7 & 29.7 & 187.8\\
\enddata
\end{deluxetable}

\end{document}